\documentclass[preprint,aps,draft,showpacs,showkeys]{revtex4}

\begin{document}
\preprint{APS/123-QED}

\title{Biosupersymmetry}

\author{Marcin Molski}

\affiliation{
Department of Theoretical Chemistry, Faculty of Chemistry\\
A. Mickiewicz University of Pozna\'n,\\
ul. Grunwaldzka 6, PL 60-780 Pozna\'n, Poland}

\email{MAMOLSKI@AMU.EDU.PL}

\date{\today}

\begin{abstract}

The growth of biological systems described by the Gompertz and West-Brown-Enquist functions is considered in the framework of the space-like supersymmetric quantum mechanics. It has been shown that the supersymmetric effect of a fermion-boson conversion has a biological analogue in the phenomenon of a growth-regression transformation under the influence of a cycle-non-specific drug of a constant concentration. The results obtained reveal that the biological growth can be viewed as the macroscopic quantum phenomenon endowed with the space-like supersymmetric properties not established so far in the domain of biology and medicine. 

\end{abstract}



\maketitle

\section{Introduction}

Supersymmetry introduced in physics by Gel'fand and Likhtman \cite{GL} is a new kind of symmetry under which 
particles possessing integer spin (bosons) and half-integer spin (fermions) can be transformed to each other. 
If supersymmetry holds, the laws of particle physics will be unaffected by this special interchange of bosons 
and fermions. Because bosons are carriers of fundamental interactions, whereas fermions are fundamental 
constituents of  matter, supersymmetry may explain the relation between forces acting in nature and its
structural component - matter. In a supersymmetric world, the concept of space–time is extended to the 
superspace being an extension of ordinary four dimensional space–time in which new dimensions appear. 
The superspace forms the background for formulation of the Standard Model in which unification of fundamental 
forces is possible \cite{Novaes}. As a result, the Standard Model gives a successful description of particle interactions that has 
been verified experimentally with meticulous precision. The Standard Model permits not only the unification of all forces but also the identification of the fundamental symmetry principles that determine the laws of nature including supersymmetry as a key component.

The concept of supersymmetry  plays an important role also in nuclear, atomic and statistical physics \cite{FCooper}. One of the useful applications of the supersymmetry in the domain of time-like quantum mechanics (SUSYQM) is obtaining analytical eigenvalues and eigenfunctions of the Schr\"odinger equation including the potential energy term for which the superpotential can be determined \cite{FCooper,Witten,Sukumar,Filho,Fukui, Molnar,Schwabl}. The first objective of the present work is extension of the time-like  
SUSYQM to include space-like quantum states, which are solution of the space-like counterpart of the Schr\"odinger equation    
\begin{equation}
-\frac{\hbar^2}{2mc^2}\frac{d^2}{d t^2}\phi(t)_v+V(t)\phi(t)_v=cP_v\phi(t)_v
\label{6b}
\end{equation}
derived by Horodecki \cite{Horodecki} from the relativistic Feinberg equation \cite{Feinberg}. Here, $V(t)$ denotes the vector potential, $m$ is the mass of a particle, $c$ is the light velocity whereas $P_v$ is quantized momentum according to the quantum number $v=0,1,2...$. The bound states of Eq. (\ref{6b}) have not been considered yet, as they are difficult to interpret in terms of temporal vibrational motion. However, in the case of anharmonic vector potential, there are no bound states in the dissociation limit and the direction of temporal motion is consistent with the arrow of time (is not of the oscillatory type). In such circumstances the space-like solutions of Eq. (\ref{6b}) can be employed to test their relevance in different areas of the science including physics, biology and medicine \cite{Molski1}. 

Another hint of supersymmetry is a discovery that the genetic coding assignments have a compact description in terms of supersymmetry \cite{Bashford,Bashford1}. In this approach the structure and evolution of the genetic code is described  
in terms of classical superalgebras employing the decomposition of a family of 64-dimensional, irreducible - typical $A(5,0)\cdot
sl(6y1)$ - representations. They evidently match the classification of codons in terms of exchange symmetries in codon
quartets. This supersymmetric model can be applied in description of the structure and evolution of the eukaryotic and 
vertebrate mitochondrial genetic codes, based on the representation theory of the Lie, in which a key role is played by 
pyrimidine and purine exchange symmetries in codon quartets.

The supersymmetry appearing in a biological context is remarkable, but still somewhat mysterious. Unless it is a weird coincidence, it points to a deep link between the quantum realm of particle physics and the quasi-classical or 
quasi-quantum realm of biological systems. The notion quasi-quantum refers to the macroscopic systems which are described in the quantum formalism including microscopic phenomena as quantization, uncertainty, coherence, nonlocality etc. In view of this 
the second objective of the present work is to prove that the concept of space-like SUSYQM is consistent with the Gompertz and West-Brown-Enquist models of growth of biological systems (organism, organ, tissue, tumor, bacterial colony), described in terms of the time-dependent growth and regression (decay) functions. The results obtained reveal that the biological growth can be viewed as the macroscopic quantum phenomenon endowed with the quasi-quantum supersymmetric properties not established so far in the life sciences. 

\section{Space-like quantum supersymmetry}

In SUSYQM it is assumed that the quantal system is characterized by the two-component Hamiltonian \cite{FCooper}
\begin{equation}
\hat{H}=\left[ 
\begin{array}{cc}
\hat{H}_-&0 \\
0 & \hat{H}_+ \\
\end{array} \right]
\label{eq1}
\end{equation}
in which
\begin{equation}
\hat{H}_-=\hat{A}^{\dag}\hat{A},\hskip1cm
\hat{H}_+=\hat{A} \hat{A}^{\dag} 
\label{eq2}
\end{equation}
represent the fermionic and bosonic components, respectively. The Hamiltonians 
$\hat{H}_-$ and $\hat{H}_+$ are said to be supersymmetric partners of each other. In the space-like version of SUSYQM, we assume that the fermionic and bosonic components $H_{\pm}$ correspond to an isospectral pair of time-dependent potentials $V(t)_{\pm}$ defined in a such manner that the ground state eigenvalue $P_0$ of the Feinberg-Horodecki equation (\ref{6b}) ($\hbar=m=c=1$) 
\begin{equation}
\left[-\frac{1}{2}\frac{d^2}{dt^2}+V(t)_{\pm}-P_0\right]\phi^{\pm}_v=\Delta P_{v0}\phi^{\pm}_v,
\hskip1cm \Delta P_{v0}=P_v-P_0
\label{eq4}
\end{equation}
is incorporated in the potentials $V(t)_{\pm}$ or is equal to zero. 

The Hamiltonians $H_{\pm}$ may be factorized into products of Hermitian conjugate operators $\hat{A}^{\dag}$ and $\hat{A}$ \cite{FCooper}
\begin{eqnarray}
\hat{H}_-=\frac{1}{\sqrt{2}}\left[-\frac{d}{dt}+W(t)\right]
\frac{1}{\sqrt{2}}\left[\frac{d}{dt}+W(t)\right]=\hat{A}^{\dag}\hat{A}, \nonumber\\
\hat{H}_+=\frac{1}{\sqrt{2}}\left[\frac{d}{dt}+W(t)\right]
\frac{1}{\sqrt{2}}\left[-\frac{d}{dt}+W(t)\right]=\hat{A} \hat{A}^{\dag}. 
\label{eq5}
\end{eqnarray}
including the superpotential $W(t)$ to be determined by solving the Ricatti equation
\begin{equation}
V(t)_{\pm}-P_0=\frac{1}{2}\left[W^2(t)\pm \frac{dW(t)}{dt}\right]. 
\label{eq3}
\end{equation}
By analogy to the ordinary quantum time-like states, supersymmetry is said to be unbroken \cite{FCooper,Witten}, if at least one of the space-like states
\begin{equation}
\phi^{\pm}_0=\exp\left[\pm\int^tW(t')dt'\right]
\label{eq6}
\end{equation}
is a true zero mode, othervise, SUSY is said to be broken dynamically. Eq. (\ref{eq6}) rewritten to the form
\begin{equation}
\frac{1}{\phi^{\pm}_0}\frac{d\phi^{\pm}_0}{dt}=\pm W(t)
\label{eq6a}
\end{equation}
can be used to determine the superpotential when the trial ground state eigenfunction is known \cite{Miao}. In this way one may derive the potential and the ground state eigenvalue associated with the system employing the Ricatti equation (\ref{eq3}). 

\section{The Gompertz model}

In order to find quasi-quantum supersymmetric characteirstic of the living systems let's consider the Gompertz function \cite{Gompertz} 
\begin{equation}
G(t)=G_0\exp\left\{ \frac{b}{a}\left[1-\exp(-at)\right]\right\}=G_{\infty}\exp\left[-\frac{b}{a}\exp(-at)\right],
\label{eq7}
\end{equation}
describing the biological growth of cell, tissue, organ, organism, tumor, bacterial colony, demographic  systems etc., in which 
\begin{equation}
G_{\infty}=\lim_{t\to\infty}G(t)=G_0\exp\left({\frac{b}{a}}\right)
\label{eq8}
\end{equation}
is the upper limit  of the growth, $a$ is retardation constant; $b$ denotes the initial growth or regression  rate constant: the sign of $b$ indicates if the system grows (+) or regresses (-). The constant $G_0=G(t=0)$ stands for the initial characteristic of the system, for instance, the initial mass, volume, diameter or number of proliferating cells. 

Having introduced the Gompertz function (\ref{eq7}) it is tempting to associate with it a superpotential to be calculated by making use of Eq.(\ref{eq6a}). To simplify the calculations one may express (\ref{eq7}) in dimensionless coordinate
\begin{equation}
\tau=\frac{a(t-t_e)}{ \sqrt{2}},\hskip1cm 
t_e=\frac{1}{a}\ln\left(\frac{2b}{a}\right)
\label{eq9}
\end{equation}
yielding
\begin{equation}
G(\tau)=G_{\infty}\exp\left[-\frac12\exp\left(-\sqrt{2}\tau\right)\right]
\label{eq10a}
\end{equation}
and then derive the superpotential 
\begin{equation}
W(\tau)=-\frac{1}{G(\tau)}\frac{dG(\tau)}{d\tau}=-\frac{1}{\sqrt{2} }\exp\left(-\sqrt{2}\tau\right)
\label{eq10}
\end{equation}
associated with the Gompertz function of growth (\ref{eq10a}).

Introducing (\ref{eq10}) into the Ricatti equation (\ref{eq3}) rewritten in $\tau$ coordinate one gets
\begin{equation}
\frac{1}{2}\left[W^2(\tau)-\frac{dW(\tau)}{d\tau}\right]=\frac{1}{4}\left[1-\exp({-
\sqrt{2}\tau})\right]^2-\frac{1}{4}=V(\tau)_{-}-P_0. 
\label{eq11}
\end{equation}
Hence, one may prove that the Gompertzian growth is governed by the second-order fermionic equation 
\begin{equation}
\hat{H}_-G(\tau)=\hat{A}^{\dag}\hat{A}G(\tau)=0
\longleftrightarrow-\frac12\frac{d^2G(\tau)}{d\tau^2}+\frac{1}{4}\left[1-\exp({-
\sqrt{2}\tau})\right]^2G(\tau)=\frac{1}{4}G(\tau),
\label{eq11a}
\end{equation}
in which 
\begin{equation}
\hat{A}=\frac{1}{\sqrt{2}}\left[\frac{d}{d\tau}-\frac{1}{\sqrt{2}}\exp(-\sqrt{2}\tau)\right],\hskip1cm
\hat{A}^{\dag}=\frac{1}{\sqrt{2}}\left[-\frac{d}{d\tau}-\frac{1}{\sqrt{2}}\exp(-\sqrt{2}\tau)\right].
\label{eq14}
\end{equation}
Eq.(\ref{eq11a}) was recently derived by Molski and Konarski \cite{Molski} by direct mathematical operations on the Gompertz function (twice differentiated with respect to the time-coordinate). One may prove (see Appendix) that Eq.(\ref{eq11a}) is the special case of the quantal Feinberg-Horodecki equation (\ref{6b}) for the time-dependent Morse oscillator \cite{Morse} 
\begin{equation}
V(\tau)_-=\frac{1}{4}\left[1-\exp({-\sqrt{2}\tau})\right]^2.
\label{eq14a}
\end{equation}
Here, $\tau={a(t-t_e)}/{\sqrt{2}}$ describes a temporal displacement from the equilibrium time $t_e$, in which the potential (\ref{eq14a}) attains the minimum equal to zero.

In a similar manner, we construct the bosonic Hamiltonian $\hat{H}_+=\hat{A}\hat{A}^{\dag}$ and the associated potential $V_+$, which can be calculated from the Ricatti equation and superpotential (\ref{eq10}) providing
\begin{equation}
V_+=\frac{1}{4}\left[1+\exp(-\sqrt{2}\tau)\right]^2.
\label{eq15}
\end{equation}
It is easy to demonstrate that the solution of the second-order bosonic equation
\begin{equation}
\hat{H}_+G(\tau)^{\dag}=\hat{A}\hat{A}^{\dag}G(\tau)^{\dag}=0
\longleftrightarrow-\frac12\frac{d^2G(\tau)^{\dag}}{d\tau^2}+\frac{1}{4}\left[1+\exp({-
\sqrt{2}\tau})\right]^2G(\tau)^{\dag}=\frac{1}{4}G(\tau)^{\dag},
\label{eq15a}
\end{equation}
which includes the potential (\ref{eq15}) is the Gompertz function of regression (decay)
\begin{equation}
G(\tau)^{\dag}=G_{\infty}^{\dag}\exp\left[\frac12\exp(-\sqrt{2}\tau)\right].
\label{eq16}
\end{equation}
We conclude that the supersymmetric partner Hamiltonians $H_{\pm}$ describe the states of Gompertzian growth and regression, which are ground eigenstates of the operators $\hat{A}$ and $\hat{A}^{\dag}$. This interpretation is consistent with the fact that the equations $\hat{A}G(\tau)=0$ and $\hat{A}^{\dag}G(\tau)^{\dag}=0$ expressed in the original time coordinate
\begin{equation}
\hat{A}G(\tau)=0\longleftrightarrow \left[\frac{d}{dt}-b\exp(-at)\right]G(t)=0,
\label{eq17}
\end{equation}
\begin{equation}
\hat{A}^{\dag}G(\tau)^{\dag}=0\longleftrightarrow \left[\frac{d}{dt}+b\exp(-at)\right]G(t)^{\dag}=0
\label{eq18}
\end{equation}
represent the well-know in biological and medical sciences the first-order kinetics of the Gompertzian growth and regression. On the other hand, they are 
a special case of the more general quantal equations describing the minimum uncertainty coherent states of the time-dependent Morse oscillator characterized by the anharmonic
constant equal to one (see Appendix). In this picture $\hat{A}$ and  $\hat{A}^{\dag}$ are interpreted as annihilation and creation operators, respectively. 

\section{The growth-regression and fermion-boson conversion}

One of the intriguing predictions of the SUSUY in the elementary particle physics is a possibility of conversion
of fermions into bosons and vice versa. For example, the supersymmetric partner of the electron is selectron, 
which can be created from the electron by absorption of the photino - supersymmetric fermionic partner 
(spin $s=1/2$) of  the bosonic photon (spin $s=1$): electron+ photino = selectron. One may prove that this
particle conversion has an analogon in the biological domain as the Gompertz growth function can be converted 
into the Gompertz function of regression. To this aim let's consider Eq. (\ref{eq17}) given in the form
\begin{equation}
\frac{1}{G(t)}\frac{dG(t)}{dt}=-a\ln\left[\frac{G(t)}{G_{\infty}}\right].
\label{eq19}
\end{equation}
If the living system, for example tumor, is exposed to a cycle-non-specific drug of a constant concentration $c$, 
which kills the cells, then the dose-response relationship between the drug concentration and the rate of the cell 
destruction is described by the equation \cite{Wheldon}
\begin{equation}  
\frac{1}{G(t)_c}\frac{dG(t)_c}{dt}=-a\ln\left[\frac{G(t)_c}{G_{\infty}}\right]-c.
\label{eq20}
\end{equation}
The solution of this equation for $b>c$ is the Gompertz function of growth 
\begin{equation}
G(t)_c=G_0\exp\left\{\left(\frac{b-c}{a}\right)\left[1-\exp(-at)\right]\right\},
\label{eq21}
\end{equation}
which for $c=2b$ undergoes conversion to the  function of Gompertzian regression. In the terms of SUSY, it can 
be interpreted as the fermionic (growth) state conversion into bosonic (regression) one under the influence of the 
external chemical field identified with the presence of drug molecules of the concentration $c>b$.

Expressing (\ref{eq21}) in $c$-dependent dimensionless coordinate
\begin{equation}
\tau_c=\frac{a(t-t_e^c)}{ \sqrt{2}},\hskip1cm 
t_e^c=\frac{1}{a}\ln\left[\frac{2(b-c)}{a}\right]
\label{eq21a}
\end{equation}
one gets the Gompertz function
\begin{equation}
G(\tau_c)=G_{\infty}\exp\left[-\frac12\exp\left(-\sqrt{2}\tau_c\right)\right]
\label{eq21b}
\end{equation}
and then the superpotential 
\begin{equation}
W(\tau_c)=-\frac{1}{G(\tau_c)}\frac{dG(\tau_c)}{d\tau_c}=-\frac{1}{\sqrt{2} }\exp\left(-\sqrt{2}\tau_c\right)
\label{eq21c}
\end{equation}
for the Gompertz function (\ref{eq21}). Taking into account the Ricatti equation (\ref{eq3}) one may calculate the fermionic potential for the $c$-treated Gompertzian system
\begin{equation}
V^c_-=\frac{1}{4}\left[1-\exp(-\sqrt{2}\tau_c)\right]^2,
\label{eq21d}
\end{equation}
which reveals that the dissociation energy of the system $D=1/4$ is unchanged under the influence of the $c$-field, whereas the equilibrium time $t_e^c$ diminishes with increasing drug's concentration $c$. 

\section{The ontogenic growth}

In 2001 West, Brown and Enquist (WBE) \cite{WBE} formulated a general model for ontogenic 
growth from the first principles. On the basis of the conservation of metabolic energy, the allometric scaling of metabolic rate, and energetic costs of producing and maintaining biomas, they derived the first-order equation  
\begin{equation}
\frac{dm}{dt}=n_1m^{p}-n_2m,\hskip1cm p=\frac34
\label{eq22}
\end{equation}
in which $n_1=B_0m_c/E_c$ and $n_2=B_c/E_c$, whereas  
$B_0$ is the initial ($t=0$) average resting metabolic rate of the whole organism, $B_c$ is the metabolic rate 
of a single cell, $E_c$ is the metabolic energy required to form a cell, $m_c$ is the mass of a cell.     

The solution of Eq.(\ref{eq22}) is the WBE function 
\begin{equation}
y(t)=\left[1-c_3\exp(-c_1t)\right]^{\frac{1}{c_2}}
\label{eq23}
\end{equation}
in which
\begin{equation}
y(t)=\frac{m(t)}{M},\hskip.5cm c_3=1-\left(\frac{m_0}{M}\right)^{c_2},\hskip.5cm 
c_1=\frac{n_1}{4M^{c_2}},\hskip.5cm  c_2=1-p.
\label{eq24}
\end{equation}
Here, $m_0=m(t=0)$ is the initial mass of the system whereas $M=m(t\to\infty)=(n_1/n_2)^{1/c_2}$ is the 
maximum body size reached. The WBE function (\ref{eq23}) fits very well the data for a variety of different species from protozoa to mammalians, and parameters fitted can be related to the biological characteristics of the system under consideration. 

Taking into account Eq. (\ref{eq6a}) one may derive the superpotential associated with the WBE growth function (\ref{eq23})  
\begin{equation}
W(t)=-\frac{1}{y(t)}\frac{dy(t)}{dt}=-\frac{\frac{c_3c_1}{c_2}\exp(-c_1t)}
{1-c_3\exp(-c_1t)},
\label{eq25}
\end{equation}
which satisfies the relation 
\begin{equation}
\frac{dW(t)}{dt}
=\frac{\frac{c_3c_1^2}{c_2}\exp(-c_1t)}{[1-c_3\exp(-c_1t)]^2}
=-c_1W(t)+c_2W(t)^2
\label{eq26}
\end{equation}
useful in solving the Ricatti equation (\ref{eq3}). Application of (\ref{eq26}) and (\ref{eq3}) provides   
\begin{equation}
\frac{1}{2}\left[W^2(t)+c_1W(t)-c_2W(t)^2\right]=
\frac{c_1^2}{8(1-c_2)}\left[\frac{2(1-c_2)}{c_1}W(t)+1\right]^2-\frac{c_1^2}{8(1-c_2)}=V(t)_- -P^-_0. 
\label{eq27}
\end{equation}
Substituting the superpotential (\ref{eq25}) into (\ref{eq27}) one gets the potential $V(t)_-$ and the ground state momentum $P^-_0$
\begin{equation}
V(t)_-=D\left\{\frac{1-\exp[-c_1(t-t_e)]}{1-s\exp[-c_1(t-t_e)]}\right\}^2,\hskip1cm P_0=D=\frac{c_1^2}{8(1-c_2)} 
\label{eq28}
\end{equation}
in which
\begin{equation}
t_e=\frac{1}{c_1}\ln\left[\frac{c_3(2-c_2)}{c_2}\right],\hskip1cm s=\frac{c_2}{2-c_2}.
\label{eq29}
\end{equation}
The derived Eq. (\ref{eq28}) respresents the well-known Wei Hua \cite{Wei} potential applied in description of anharmonic vibrations of diatomic molecules. Here $D$ represents the dissociation energy of the system, whereas $t_e$ is an equilibrium time point at which potential attains minimum equal to zero $V_-(t=t_e)=0$. 

Employing the superpotential $W(t)$ associated with the WBE function of growth,
one may construct the fermionic operator $\hat{H}_-=\hat{A}^{\dag}\hat{A}$ in which 
\begin{equation}
\hat{A}=\frac{1}{\sqrt{2}}\left[\frac{d}{dt}-\frac{\frac{c_3c_1}{c_2}\exp(-c_1t)}
{1-c_3\exp(-c_1t)}\right],\hskip1cm
\hat{A}^{\dag}=\frac{1}{\sqrt{2}}\left[-\frac{d}{dt}-\frac{\frac{c_3c_1}{c_2}\exp(-c_1t)}
{1-c_3\exp(-c_1t)}\right],
\label{eq30}
\end{equation}
and then the the second-order fermionic equation 
\begin{equation}
\left[-\frac{1}{2}\frac{d^2}{dt^2}+V(t)_{-}-P_0\right]y(t)=0,
\label{eq29a}
\end{equation}
which represents the quantal Horodecki-Feinberg equation ($\hbar=m=c=1$) for the Wei Hua time-dependent oscillator in the dissociation state in which momentum eigenvalue is equal to the dissociation energy of the system (see Appendix). 

In the similar manner one may construct the bosonic operator $\hat{H}_+=\hat{A}\hat{A}^{\dag}$ associated with the potential $V_+$ to be calculated from the Ricatti equation 
\begin{equation}
\frac{1}{2}\left[W^2(t)-c_1W(t)+c_2W(t)^2\right]=
\frac{c_1^2}{8(1+c_2)}\left[\frac{2(1+c_2)}{c_1}W(t)+1\right]^2-\frac{c_1^2}{8(1+c_2)}=V(t)_+ -P'_0, 
\label{eq30a}
\end{equation}
in which
\begin{equation}
V(t)_+=D'\left\{\frac{1+\exp[-c_1(t-t'_e)]}{1-s'\exp[-c_1(t-t'_e)]}\right\}^2,
\hskip1cm D'=P'_0=\frac{c_1^2}{8(1+c_2)},
\label{eq31}
\end{equation}
\begin{equation}
t'_e=\frac{1}{c_1}\ln\left[\frac{c_3(2+c_2)}{c_2}\right],\hskip1cm s'=\frac{c_2}{2+c_2}.
\label{eq31a}
\end{equation}
It is easily to prove that the eigenfunction of the bosonic operator $\hat{H}_+=\hat{A}\hat{A}^{\dag}$ is the WBE function of regression
\begin{equation}
y(t)^{\dag}=\left\{1-s'\exp-c_1(t-t'_e)]\right\}^{-{\frac{1}{c_2}}}=
\left\{1-s\exp-c_1(t-t_e)]\right\}^{-{\frac{1}{c_2}}}=
\left[1-c_3\exp(-c_1t)\right]^{-{\frac{1}{c_2}}},
\label{eq33}
\end{equation}
which has not been considered so far in the medical and biological domains. It is esay to prove that the transition from the set of growth to regression parameters can be realized under the parameter transformation $c_2\to -c_2$ yielding
\begin{equation}
\left(c_2,V_-,D,t_e,s\right)\longrightarrow \left(-c_2,V_+,D',t'_e,s'\right).
\label{eq33a}
\end{equation}
Having introduced the operators (\ref{eq30}) one may prove that growth and regression WBE functions satisfy  
\begin{equation}
\frac{1}{\sqrt{2}}\left\{\frac{d}{dt}-\frac{\frac{c_3c_1}{c_2}\exp(-c_1t)}
{1-c_3\exp(-c_1t)}\right\}y(t)=
\frac{1}{\sqrt{2}}\left\{\frac{d}{dt}-\frac{\frac{sc_1}{c_2}\exp[-c_1(t-t_e)]}
{1-s\exp[-c_1(t-t_e)]}\right\}y(t)=0,
\label{eq34}
\end{equation}
\begin{equation}
\frac{1}{\sqrt{2}}\left\{-\frac{d}{dt}-\frac{\frac{c_3c_1}{c_2}\exp(-c_1t)}
{1-c_3\exp(-c_1t)}\right\}y(t)^{\dag}=
\frac{1}{\sqrt{2}}\left\{-\frac{d}{dt}-\frac{\frac{sc_1}{c_2}\exp[-c_1(t-t_e)]}
{1-s\exp[-c_1(t-t_e)]}\right\}y(t)^{\dag}=0.
\label{eq34a}
\end{equation}
It is noteworthy that they are a special case of the more general quantal equations describing coherent states of the Wei Hua oscillator in the dissociation state (see Appendix). In view of the above, the operators of biological growth $\hat{A}$ and regression $\hat{A}^{\dag}$ are the special case of the quantal annihilation and creation operators appearing in the domain of quantum physics \cite{Molski2}. 

The equations derived in this section are more general then those obtained in the WBE scheme in which $p=3/4$, which corresponds to $c_2=1/4$. It is worth noticing that Eq.(\ref{eq22}) and its solution have been applied also in description of the tumor growth, in particular primary breast cancer \cite{Spratt} and in study of the tumor xenografts \cite{Michelson}.

\section{The universal growth model}

The WBE function (\ref{eq23}) can be expressed in dimensionless time-coordinate \cite{WBE}
\begin{equation}
\tau= c_1(t-t_e)\hskip1cm t_e=\frac{1}{c_1}\ln(c_3) 
\label{U1}
\end{equation}
yielding
\begin{equation}
r(\tau)=1-\exp(-\tau),
\label{U2}
\end{equation}
in which $r(\tau)=y(t)^{c_2}$. As it has been proved by WBE \cite{WBE}, the function (\ref{U2}) provides the powerful way of plotting the data that reveals universal properties of biological growth. If the mass ratio $r(\tau)$ is plotted against a variable $\tau$ then all species (mammals, birds, fish, crustacea), regardless of taxon, cellular metabolic rate $B_c$ and mature body size $M$ fall on the same parameterless universal curve (\ref{U2}).   

Taking advantage of Eq. (\ref{eq6a}) one may derive the superpotential $W(\tau)$ and then potential $V(\tau)_-$ and momentum $P_0$ associated with the universal growth function (\ref{U2})
\begin{equation}
W(\tau)=-\frac{\exp(-\tau)}{1-\exp(-\tau)},
\label{U3}
\end{equation}
\begin{equation}
\frac{1}{2}\left[W(\tau)^2-\frac{dW(\eta)}{d\eta}\right]=
V(\tau)_--P_0=-\frac{1}{2}\frac{\exp(-\tau)}{1-\exp(-\tau)}.
\label{U4}
\end{equation}
Having introduced the superpotential $W(\tau)$,one may construct the fermionic Hamiltonian $\hat{H}_-=\hat{A}^{\dag}\hat{A}$ in which 
\begin{equation}
\hat{A}=\frac{1}{\sqrt{2}}\left[\frac{d}{d\tau}-\frac{\exp(-\tau)}{1-\exp(-\tau)}\right],\hskip1cm
\hat{A}^{\dag}=\frac{1}{\sqrt{2}}\left[-\frac{d}{d\tau}-\frac{\exp(-\tau)}{1-\exp(-\tau)}
\right],
\label{U5}
\end{equation}
and then the the second-order equation 
\begin{equation}
\left[-\frac{1}{2}\frac{d^2}{d\tau^2}+V(\tau)_{-}-P_0\right]r(\tau)=0,
\label{U6}
\end{equation}
which represents the quantal Horodecki-Feinberg equation ($\hbar=m=c=1$) for the  time-dependent Hulth\'{e}n potential \cite{Hulthen} and {\it critical screening} 
\cite{Varshni} leading to $P_0=0$ (see Appendix).   

In the similar manner one may construct the bosonic Hamiltonian $\hat{H}_+=\hat{A}\hat{A}^{\dag}$ associated with the potential $V_+$ to be calculated from the Ricatti equation 
\begin{equation}
\frac{1}{2}\left[W(\tau)^2+\frac{dW(\eta)}{d\eta}\right]=
V(\tau)_+-P_0=\frac{1}{2}\frac{\exp(-2\tau)+\exp(-\tau)}{[1-\exp(-\tau)]^2}.
\label{U7}
\end{equation}
The eigenfunction of the bosonic operator $\hat{H}_+=\hat{A}\hat{A}^{\dag}$ is the WBE universal function of regression
\begin{equation}
r(\tau)^{\dag}=\left[1-\exp(-\tau)\right]^{-1}.
\label{U8}
\end{equation}
One may prove that growth and regression universal functions satisfy the first-order differential equations   
\begin{equation}
\frac{1}{\sqrt{2}}\left[\frac{d}{d\tau}-\frac{\exp(-\tau)}{1-\exp(-\tau)}\right]r(\tau)=0,
\hskip1cm
\frac{1}{\sqrt{2}}\left[-\frac{d}{d\tau}-\frac{\exp(-\tau)}{1-\exp(-\tau)}
\right]r(\tau))^{\dag}=0.
\label{U9}
\end{equation}
 
\section{Conclusions}

The results obtained demonstrate a possibility of description of the biological systems in terms of the space-like SUSYQM. In particular the growth (regression) function can be interpreted as the fermion (boson) solution of the partner Hamiltonians, which form a complete set of eigenfunctions satisfying
\begin{equation}
\left[ 
\begin{array}{ll}
\hat{H}_-&0 \\
0 & \hat{H}_+ \\
\end{array} \right]
\left[ 
\begin{array}{l}
G(t) \\
G(t)^{\dag} \\
\end{array} \right]=0.
\label{eq111}
\end{equation}
The Gompertz growth and regression functions can be converted to each other under the influence of the external chemical field generating the dose-response relationship between drug concentration and rate of the cell destruction. Associating with the Gompertzian growth the superpotential one may derive - employing the Riccati equation - the corresponding potential and introduce the second-order kinetic equation of growth complementary to that widely applied in medical and biological sciences the first-order Gompertzian kinetics. The derived Morse potential is well-known in the domain of molecular anharmonic oscillators whereas the second-order kinetic equation is the special case of the space-like Feinberg-Horodecki quantal equation for the time-dependent Morse oscillator in the dissociation state. It means that the momentum eigenvalue (in units $c=1$) is equal to the dissociation energy of the system. In such circumstances the direction of temporal motion is consistent with the arrow of time - it is not of the oscillatory type. The biosupersymmetric approach can be applied also to the ontogenic growth described by the WBE function, which represents the ground-state solution of the Feinberg-Horodecki quantal equation for the time-dependent Wei Hua oscillator in the dissociation state.  

The results obtained reveal an unknown connection between the quantum realm of particle physics - represented by SUSYQM, and the macroscopic biological systems, which can be described in the quantum formalism employing such notions as quantization, uncertainty, coherence and nonlocality. In particular, the macroscopic Gompertz and WBE functions are the solution of the quantal eigenvalue equation, with discrete eigenvalue equal to the dissociation energy of the system. This equation represents the second-order Gompertzian
and WBE kinetics so far not applied in the domain of the biological and medical sciences. 
The Gompertz and WBE functions are also the ground-state solutions of the first-order equation, which is the special case of the  quantal formulae describing space-like minimum-uncertainty coherent states of the time-dependent Morse and Wei Hua oscillators. In particular one may prove that the macroscopic Gompertz and WBE growth functions  represent the so-called {\it intelligent} coherent states \cite{Aragone}, which not only minimize the time-energy uncertainty relation but also maintain this relation in space due to its spatial stability \cite{Molski1}. Such space-like states differ from the ordinary time-like coherent states, which maintain the position-momentum Heisenberg relation in time due to its temporal stability \cite{Zhang, Cooper}.

Although the concepts of supersymmetry, quantization, uncertainty, coherence and nonlocality are restricted to the area of quantum physics, the results obtained reveal existence of a new class of macroscopic quantum phenomena, which play an important role in the biological domain. According to the Leggett's classification \cite{Leggett}, one may distinguish the macroscopic quantum phenomena of the first kind (superfluidity, superconductivity) and the second kind (quantum interference of macroscopically distinct states, macroscopic tunneling via Josephson junction). The results obtained justify the introduction in the Leggett's classification of a new class of the quasi-quantum phenomena: macroscopic quantization, uncertainty, coherence and nonlocality appearing in the Gompertzian and WBE systems. In such circumstances, the biological growth can be viewed as the macroscopic quantum phenomenon of the third kind.  

\section*{Appendix}

\subsection*{The Morse oscillator}

One may prove that Eq. (\ref{eq11a}) is the special case of the more general quantal Feinberg-Horodecki equation (\ref{6b}) \cite{Molski1} 
\begin{equation}
-\frac{\hbar^2}{2mc^2}\frac{d^2}{dt^2}\phi(t)_v+V(t)\phi(t)_v=cP_v\phi(t)_v
\label{M1}
\end{equation}
for the time-dependent Morse oscillator characterized by the potential
\begin{equation}
V(t)=D\left\{1-\exp[-a(t-t_e)]\right\}^2.
\label{M2}
\end{equation}
Expressing (\ref{M1}) in the dimensionless coordinate $\tau=a(t-t_e)(2x_e)^{-1/2}$
one gets
\begin{equation}
-\frac12\frac{d^2\phi(\tau)_v}{d\tau^2}+\frac{1}{4x_e}\left[1-\exp({-
\sqrt{2x_e}\tau})\right]^2\phi(\tau)_v=\frac{P_v}{\hbar(\omega_e/c)}\phi(\tau)_v
\label{M3}
\end{equation}
in which 
\begin{equation}
P_v=\hbar(\omega_e/c)\left[\left(v+\frac12\right)-
\left(v+\frac12\right)^2x_e\right],\hskip1cm v=0,1,2.......
\label{M4}
\end{equation}
is the quantized momentum eigenvalue; $\omega_e=(a/c)(2D/m)^{1/2}$ is the vibrational frequency defined by the reduced mass $m$ of the oscillator, whereas $x_e$ is the anharmonicity constant $x_e={\hbar\omega_e}/(4D)$ defined by the Planck's constant $h=\hbar 2\pi$. It is easy to prove that for $x_e=1$ and $v=0$, the quantal Eq. (\ref{M3}) transforms to macroscopic Eq. (\ref{eq11a}) whereas the ground state eigenfunction $\phi(\tau)_0$ takes the identical form as the Gompertz growth function $G(\tau)$ with the accuracy to the multiplicative constant $G_{\infty}$
\begin{eqnarray}
\exp\left[-\frac{1}{2x_e}\exp\left(-\sqrt{2x_e}\tau\right)\right]
\exp\left[-\frac{1}{\sqrt{2x_e}}(1-x_e)\tau\right]
&\to&\exp\left[-\frac12\exp\left(-\sqrt{2}\tau\right)\right]\nonumber\\
\psi_0(\tau)&\to &G(\tau)
\label{M5}
\end{eqnarray}
The condition $x_e={\hbar\omega_e}/(4D)=1$ corresponds to the relationship $P_0c=\hbar\omega_e/4=D$ indicating that the Morse oscillator is in dissociation state,
which is characterized by the nonvanishing momentum $P_0=\hbar\omega_e/(4c)$. Hence, the time-evolution of the Gompertzian systems is consistent with the arrow of time.
The same results one gets from (\ref{M4}) by substituting $v=0$ and $x_e=1$. 

One may prove that the Gompertzian equations of growth and regression are 
a special case of the more general quantal equations describing the minimum uncertainty coherent states of the time-dependent Morse oscillator characterized by the anharmonic
constant equal to one. To demonstrate this property, the Feinberg-Horodecki equation (\ref{eq11b}) is expressed in terms of the annihilation and creation operators  $\hat{A}$ and  $\hat{A}^{\dag}$ \cite{Molski,Molski1}
\begin{equation}
\hat{A}^{\dag}\hat{A}|v>=v[1-(v+1)x_e]|v>,
\label{M6}
\end{equation}
defined in the following manner
\begin{equation}
\hat{A}= \frac{1}{\sqrt{2}}\left[\frac{d}{d \tau}+ \frac{1}{\sqrt{2x_e}}\left[1-\exp({-\sqrt{2x_e}\tau})\right]-\sqrt{\frac{x_e}{2}}\right]
\label{M7}
\end{equation}
\begin{equation}
\hat{A}^{\dag}= \frac{1}{\sqrt{2}}\left[-\frac{d}{d \tau}+ \frac{1}{\sqrt{2x_e}}\left[1-\exp({-\sqrt{2x_e}\tau}\right)-\sqrt{\frac{x_e}{2}}\right].
\label{M8}
\end{equation}
The space-like coherent states of the Morse oscillator are eigenstates of the annihilation operator
\begin{equation}
\hat{A}|\alpha>=\alpha|\alpha>
\label{II.50}
\end{equation}
\begin{equation}
|\alpha>=\exp(\sqrt{2\alpha}\tau)\exp\left[-\frac{1}{2x_e}
\exp(-\sqrt{2x_e}\tau)\right]\exp\left[-\frac{1}{\sqrt{2x_e}}(1-x_e)\tau\right].
\label{M10}
\end{equation}
Such states minimize the time-energy uncertainty relation \cite{Molski, Molski1} 
\begin{equation}
(\Delta T(\tau))^2(\Delta E)^2 \ge\frac14{<\alpha|g(\tau)|\alpha>^2},\hskip.5cm
[T(\tau),\hat{E}]=ig(\tau)=i\exp\left({-{\sqrt{2x_e}}\tau}\right),
\label{M11}
\end{equation}
yielding
\begin{equation}
(\Delta T(\tau))^2(\Delta E)^2 =\frac14{<\alpha|g(\tau)|\alpha>^2}
\label{M12}
\end{equation}
in which 
\begin{equation}
(\Delta T(\tau))^2=<\alpha|T(\tau)^2|\alpha> - <\alpha|T(\tau)|\alpha>^2,\hskip.2cm
(\Delta E)^2=<\alpha|\hat{E}^2|\alpha> - <\alpha|\hat{E}|\alpha>^2,
\label{M13}
\end{equation}
\begin{equation}
\hat{E}=\frac{d}{d\tau}\hskip.2cm (\hbar=1),\hskip.5cm T(\tau)=\frac{1}{\sqrt{2x_e}}\exp\left({-\sqrt{2x_e}\tau}\right) + \frac{1}{\sqrt{2x_e}}(1-x_e),                 
\label{M14}
\end{equation}
whereas $|\alpha>$ is given by (\ref{M10}). In the above formulae, $T(\tau)$ denotes, to within a constant, the temporal dimensionless Morse variable. 

If we introduce $x_e=1$ into Eqs.(\ref{M7})-(\ref{M14}) and restrict considerations to the ground coherent states $|0>$ and $<0|$ with $\alpha=0$, one may prove that 
\begin{eqnarray}
\frac{1}{\sqrt{2}}\left\{\frac{d}{d \tau}+ \frac{1}{\sqrt{2x_e}}\left[1-\exp({-\sqrt{2x_e}\tau})\right]-\sqrt{\frac{x_e}{2}}\right\}|0>
\longrightarrow&\nonumber\\
\frac{1}{\sqrt{2}}\left[\frac{d}{d\tau}-\frac{1}{\sqrt{2}}\exp(-\sqrt{2}\tau)\right]G(\tau)=
\left[\frac{d}{dt}-b\exp(-at)\right]G(t)=0,&
\label{M15}
\end{eqnarray}

\begin{eqnarray}
<0|\frac{1}{\sqrt{2}}\left\{-\frac{d}{d \tau}+ \frac{1}{\sqrt{2x_e}}\left[1-\exp({-\sqrt{2x_e}\tau}\right)-\sqrt{\frac{x_e}{2}}\right\}
\longrightarrow&\nonumber\\
\frac{1}{\sqrt{2}}\left[-\frac{d}{d\tau}-\frac{1}{\sqrt{2}}\exp(-\sqrt{2}\tau)\right]G(\tau)^{\dag}=
\left[\frac{d}{dt}+b\exp(-at)\right]G(t)^{\dag}=0.&
\label{M16}
\end{eqnarray}

\begin{equation}
(\Delta T(\tau))^2(\Delta E)^2 =\frac14{<0|g(\tau)|0>^2},\hskip.2cm g(\tau)=\exp\left({-{\sqrt{2}}\tau}\right)\hskip.2cm
T(\tau)=\frac{1}{\sqrt{2}}\exp\left({-\sqrt{2}\tau}\right)
\label{M17}
\end{equation}

The results obtained above can be summarized in the following points.
\begin{description}
\item[i.] The second-order macroscopic equation governing the Gompertzian growth is a special case of the space-like quantum Horodecki-Feinberg equation for the time-dependent Morse oscillator whose ground state $v=0$ eigenfunction for $x_e=1$ reduces to the macroscopic Gompertz function of growth.

\item[ii.] The macroscopic second-order Gompertzian kinetics is described by the quasi-quantum equation whose eigenvalue is quantized and takes only one value $1/4$
equal to the dimensionless dissociation energy of the Morse oscillator, hence the latter is in the dissociation state - there is no oscillation in time and evolution of the system is consistent with the arrow of time.

\item[iii.] The first-order macroscopic equations governing the Gompertzian growth and regression are the special case of the space-like quantum annihilation and creation equations. For $\alpha=0$ the ground state eigenfunctions of the annihilation and creation operators, reduce to the the Gompertz function of growth and regression. 

\item[iv.] The Gompertz function of growth has an identical form as the minimum uncertainty coherent state $\alpha=0$ of the time-dependent Morse oscillator characterized by the anharmonic constant equal to one. Such states minimize the time-energy uncertainty relation and maintain this relation in space.  
 
\end{description}

\subsection*{The Wei Hua oscillator}

In the case of the time-dependent Wei Hua oscillator \cite{Wei}, the Feinberg-Horodecki equation (\ref{6b}) takes the form 
\begin{equation}
-\frac{\hbar^2}{2mc^2}\frac{d^2}{d t^2}\phi(t)_v+
D\left\{\frac{1-\exp[-c_1(t-t_e)]}{1-s\exp[-c_1(t-t_e)]}\right\}^2
\phi(t)_v=cP_v\phi(t)_v.
\label{W1}
\end{equation}
The ground state ($v=0)$ eigenfunction of (\ref{W1}) can be written in the form \cite{Wei}
\begin{equation}
\phi(t)_0=\left\{1-s\exp[-c_1(t-t_e)]\right\}^{\frac{1}{c_2}}\left\{s\exp[-c_1(t-t_e)]
\right\}^{c_0},
\label{W2}
\end{equation} 
in which \cite{Wei}
\begin{equation}
c_0=\left[\frac{2mc^2}{\hbar^2c_1^2}(D-cP_v)\right]^{1/2},\hskip1cm \frac{1}{c_2}=\left[\frac{2mc^2D}{\hbar^2c_1^2}\right(\frac{1}{s}-1\left)+\frac14\right]^{1/2}+\frac12,
\label{Eq37b}
\end{equation}
The annihilation and creation operators as well as the minimum-uncertainty coherent states of the Wei Hua oscillator, which satisfy equation (\ref{II.50}) can be given in the form   
\begin{equation}
\hat{A}=\frac{1}{\sqrt{2}}\left[\frac{d}{dt}
-\frac{(c_1s/c_2)\exp[-c_1(t-t_e)]}{1-s\exp[-c_1(t-t_e)]}+c_1c_0\right],
\label{Eq37f}
\end{equation}
\begin{equation}
\hat{A}^{\dag}= \frac{1}{\sqrt{2}}\left[-\frac{d}{d t}
-\frac{(c_1s/c_2)\exp[-c_1(t-t_e)]}{1-s\exp[-c_1(t-t_e)]}+c_1c_0\right],
\label{Eq37g}
\end{equation}
\begin{equation}
|\alpha>=\left\{1-s\exp[-c_1(t-t_e)]\right\}^{\frac{1}{c_2}}\left\{s\exp[-c_1(t-t_e)]
\right\}^{c_0}
\exp(\sqrt{2}\alpha t).
\label{Eq37h}
\end{equation} 
In the ground coherent state $\alpha=0$ and the dissociation limit $D=cP_0$ we have $c_0=0$, hence the quantal function (\ref{Eq37h}) reduces to the WBE function of growth
\begin{equation}
|0>=\left\{1-s\exp[-c_1(t-t_e)]\right\}^{\frac{1}{c_2}} = 
\left[1-c_3\exp(-c_1t)\right]^{\frac{1}{c_2}}=y(t)
\label{Eq37i}
\end{equation}
whereas the annihilation and creation operators (\ref{Eq37f}) and (\ref{Eq37g}) take the form of the WBE growth and regression operators (\ref{eq30}).    

\subsection*{The Hulth\'{e}n oscillator}

The Feinberg-Horodecki equation (\ref{6b}) with the time-dependent Hulth\'{e}n potential
\cite{Hulthen} 
\begin{equation}
-\frac{\hbar^2}{2mc^2}\frac{d^2}{d t^2}\phi(t)_v
-V_0\frac{\exp[-c_1( t-t_e)]}{1-\exp[-c_1(t-t_e)]}\phi(t)_v=cP_v\phi(t)_v,
\label{H1}
\end{equation}
can be specified in a dimensionless form \cite{Flugge}
\begin{equation}
\left[\frac{d^2}{d\tau^2}+\beta^2\frac{\exp(-\tau)}{1-\exp(- \tau)}-\epsilon^2_v\right]\phi(\tau)_v=0,
\label{H2}
\end{equation}
in which
\begin{equation}
\tau =c_1(t-t_e),\hskip1cm \beta^2=\frac{2mc^2V_0}{\hbar^2c_1^2},\hskip1cm 
\epsilon_v^2=-\frac{2mc^3P_v}{\hbar^2c_1^2}.
\label{H3}
\end{equation}
The eigenfunctions and eigenvalues of (\ref{H2}) take the form \cite{Flugge}
\begin{equation}
\phi(\tau)_v=\exp(-\epsilon_v\tau)[1-\exp(-\tau)]_2F_1(2\epsilon_v + 1 + v, 1 - v, 2\epsilon_v + 1; \exp(-\tau))
\label{H4}
\end{equation}
\begin{equation}
\epsilon_v =\frac{\beta^2 -v^2}{2v\beta}\hskip1cm v=1,2,3.....
\label{H5}
\end{equation}
For $\beta=1$ \cite{Setare} and ground state $v=1$ we have $\epsilon^2_1=P_1=0$, whereas the eigenfunction \ref{H4}) reduces to the universal growth function (\ref{U2}) 
\begin{equation}
\phi(\tau)_1=1-\exp(-\tau)=r(\tau).
\label{H6}
\end{equation}
This result indicates that the universal growth function (\ref{U2}) can be identified with the ground state solution of the Feinberg-Horodecki equation (\ref{H1}) for the time-dependent Hulth\'{e}n oscillator at the critical screening \cite{Varshni}. Then 
$\beta=1$ and the momentum eigenvalue is equal to zero. The condition $\beta=1$ implies
$V_0={\hbar^2c_1^2}/{(2mc^2)}$, hence the Feinberg-Horodecki equation (\ref{H1}) for the critical screening can be specified in the form 
\begin{equation}
-\frac{d^2}{d t^2}\phi(t)_1
-c_1^2\frac{\exp[-c_1( t-t_e)]}{1-\exp[-c_1(t-t_e)]}\phi(t)_1=0.
\label{H7}
\end{equation}
Hence, the solution of (\ref{H7}) is the universal growth function (\ref{U2}) expressed in the original time-variable
\begin{equation}
\phi(t)_1=1-\exp[-c_1(t-t_e)].
\label{H8}
\end{equation}

\end{document}